# Tunable Thermal Expansion Behavior in the Intermetallic YbGaGe

F. R. Drymiotis[1], Y. Lee[2], G. Lawes[1], J. C. Lashley[1], T. Kimura[1], S. M. Shapiro[2], A. Migliori[1],V. Correa[1] and R. A. Fisher[3]

*[1]Los Alamos National Laboratory, Los Alamos, New Mexico 87545*
*[2]Physics Department, Brookhaven National Laboratory Upton, New York 11973*
*[3]Lawrence Berkeley National Laboratory, University of California, Berkeley, California 94720*

We investigate the effects of carbon and boron doping on the thermal expansion in the hexagonal ($P6_3/mmc$) intermetallic YbGaGe. X-ray powder diffraction was used to measure the lattice constants on pure and doped (C or B at nominal levels of 0.5 %) samples from T~10 K to T~300 K. Also measured were resistivity, specific-heat, and magnetic susceptibility. While the pure YbGaGe samples exhibit positive thermal volume expansion, $(V_{300K}-V_{10K})/V_{300K}$ = 0.94%, the volume expansion in the lightly C and B-doped samples, contract and tend towards zero volume expansion. Such a strong response with such light doping suggests that the underlying mechanism for the reported zero volume expansion is substitutional disorder, and not the previously proposed valence fluctuations.

## I. INTRODUCTION

When anisotropic thermal expansion was first reported in the hexagonal intermetallic YbGaGe, namely normal positive-thermal-expansion along the *c*-axis and slightly negative expansion along the *a*-axis, it was attractive from technological and fundamental physics viewpoints.[1] The resulting zero-volume expansion (ZVE) between 10 and 300 K was attributed to valence fluctuations between $Yb^{2+}$ and $Yb^{3+}$, a theoretical postulate advanced by Salvador and coworkers[1] and Sleight.[2] However, later work on YbGaGe failed to reproduce the results shown in Ref. 1 and showed no evidence of a $Yb^{2+} \rightarrow Yb^{3+}$ valence fluctuation.[3-5] Further work by the group in Ref.1 failed to reproduce the initially reported ZVE behavior but showed that the thermal expansion of YbGaGe is sensitive to departures ideal stoichiometry e.g. $YbGa_{1+x}Ge_{1-x}$ or $YbGa_{1-x}Ge_{1+x}$.[6] Because the YbGaGe samples reported in Ref. 1 were prepared in graphite crucibles, and those reported in Ref. 3-5 were not, it was anticipated that carbon contamination might be important; hence it could be possible to tune the thermal expansion by carbon doping. Although the thermal expansion of our stoichiometric samples show little variation with the method of preparation, the doped samples show a 47 % contraction.

## II. EXPERIMENT

YbGaGe crystallizes in the hexagonal $P6_3/mmc$ (m = 3, z = 9, no. 194 space group) structure of the MoC type. This is a layered structure consisting of puckered Ga-Ge planes sandwiched between Yb planes. There are four Ga-Ge layers per unit cell. Single crystals of YbGaGe were prepared by placing the constituents in alumina or yttria crucibles, in the molar ratio 1:1:1. Sample purities were 99.999%, 99.999% and 99.9999%, respectively. The crucibles were inserted in quartz tubes, which were evacuated, sealed, and heated in a furnace to 1000 °C, where they remained for a minimum of 24 h in order to ensure homogeneity. Following the heat treatment they were either allowed to cool slowly at a rate of 6 °C/h to room temperature or taken out of the furnace and allowed to cool rapidly. The slow-cooled profile was used for two of the samples with one of them grown in an alumina crucible and the second in a high-purity yttria crucible. The third sample was grown in an alumina crucible, but was removed from the furnace at a high temperature (~900 °C) and quenched (Q) in water. Because the samples showing ZVE in Ref 1. were prepared in graphite crucibles we hypothesized that the ZVE was disorder-driven and could be tuned by small additions of carbon. Consequently we prepared carbon and in addition boron-doped samples in alumina crucibles in molar ratios of (1:1:1:0.005) and cooled to room temperature at °6 C/h. The samples studied along with their respective preparation conditions are listed in Table 1.

Variable temperature x-ray powder diffraction measurements were measured at beamline X7A at the National Synchrotron Light Source at Brookhaven National Laboratory. Each powdered sample was loaded into a 0.3 mm glass capillary. A monochromatic beam was obtained using a channel-cut Ge (111) monochromator, and the wavelength of 0.6985(1) Å (or 0.6545(1) Å for the doped samples) was calibrated using a $CeO_2$ standard (SRM 674). The capillary was subsequently sealed and mounted on the second axis of the diffractometer inside a modified closed-cycle He cryostat, which operates under vacuum and has an absolute temperature accuracy of ±5 K and stability better than 0.1 K. Diffraction data were measured from 10 K (or 20 K for the doped samples) using a PSD stepped in 0.25° intervals over the angular range of $10° < 2\theta < 50°$ with counting times of 30 s/step. The sample temperature was increased in 50 K steps to 400 K (or 300 K for the doped samples). The capillary was rocked by $5^o$ during data collection to obtain better powder averaging. Diffraction data were fitted using LeBail method implemented in the GSAS suite of programs.[7,8] and the lattice parameters and cell volumes were determined.

We measured four-terminal resistance and specific heat in a Quantum Design physical property measurement system (PPMS) and magnetic susceptibility in a Quantum Design magnetic property measurement system (MPMS).

## III. RESULTS AND DISCUSSION

The x-ray powder diffraction pattern of YbGaGe sample slowly cooled in an yttria crucible is shown in Figure 1. An excellent fit is achieved at all temperatures with the known hexagonal symmetry of YbGaGe (P63/mmc). There are small impurity peaks that haven't been identified, but their intensity is 100 times less than the strongest Bragg peaks. The lattice parameters, *a* and *c*, of all samples measured at 300 K are given in Table 1 along with the calculated volume of the unit cell, $V = a^2 c(\sqrt{3}/2)$. Figure 2 shows the temperature dependence of the unit-cell volume and the edge lengths for the yttria grown stoichiometric sample Fig. 2 (a) and the B-doped alumina grown sample Fig. 2 (b). Figure 3 shows the relative volume change of all samples normalized to the low temperature value. From these figures it is clearly seen that normal positive thermal expansion is observed in all samples. The stoichiometric samples all display similar volume expansion whereas the C and B-doped samples have much reduced volume dependence on temperature. The volume expansion is non linear function of temperature but we can define a volume expansion coefficient by assuming a linear behavior between the volume, extrapolated to zero temperature and that measured at 300K: $\beta = \mathrm{V(300)\text{-}V(0)}/\mathrm{V(300)}\,\Delta\mathrm{T}$ , where $\Delta$T=300K. This is also shown in Table 1. The yttria grown and the alumina quenched sample have nearly the same volume expansion ($\beta$=3.2 x $10^{-5}$/K), whereas the B doped material has $\beta = 1.70$ x $10^{-5}$/K. This corresponds to a 47% decrease in thermal expansion with only 0.5% addition of impurities. The relative changes in the *c*- and *a*- axes for the yttria grown stoichiometric samples and the B-doped samples are (7.5%, 3.5%) and (2.8%, 1.1%), respectively. Both the *a*- and *c*-axes thermal expansion of YbGaGe exhibit strong dependence on very weak dopant concentration. The values of *a* and *c* lattice parameters are also dependent upon the method of growth and the impurity composition. The differences are well beyond the experimental errors, which are only errors in the last decimal place as indicated in table 1. We can conclude from these measurements that impurity composition strongly affects the thermal expansion of YbGaGe.

We now describe the resistivity and the magnetic behavior of the samples to establish the role of disorder and magnetism in the volume expansion. Figure 4a shows the temperature dependence of the residual resistance ratio (RRR), defined as *R(T)/R(300K)*, on the five samples. Of the stoichiometric samples, the yttria grown, slow-cooled sample has the largest RRR ~2.1 whereas the alumina grown slow-cooled sample and the alumina grown quenched sample have RRR values of 1.9 and 1.7 respectively. For the quenched sample, our RRR value is similar to that of the original work.[1] The resistance varies smoothly with temperature showing no evidence for a valence transition. The reduced RRR of the quenched sample is most probably caused by enhanced structural disorder scattering that could persist to high temperatures. In the case of the slow-cooled C- and B- doped samples the RRR is 1.9 and 2.2 respectively. This result is rather surprising as we expected reduced values of RRR with addition of impurities. Particularly in the case of the B doped sample, we observe the largest RRR and an effective moment comparable to the quenched stoichiometric sample. To interpret our findings, we need to emphasize the difference between structural disorder (dislocations, internal stresses etc) and substitutional disorder (substitution of Ga or Ge atoms by B and C). Although in the case of the doped samples, we see a large effect on thermal expansion we also see increased RRR, consistent with the slow-cooled growth process but not with our observed enhanced Yb effective moment. For stoichiometric samples we observed the same trends in RRR and effective moment, but we did not observe large changes in the thermal expansion behavior. This observation further substantiates the claim that the changes in the thermal expansion behavior shown above are not caused by the amount of structural disorder but rather by substitutional disorder, the substitution of Ga or Ge atoms by B and C.

We characterized the magnetic properties of all three YbGaGe samples by measuring the DC susceptibility (defined as M/H) in an external magnetic field of H=1T and made a diamagnetic correction for the sample holder, about 30% of the total signal at higher temperatures. The temperature dependent susceptibilities for all samples are shown in Figure 4b. The Yb effective moment, shown in Table 1, appears to vary with either type of disorder (substitutional or structural), and is not correlated with changes in thermal expansion. The resistance measurements show that the slow-cooled B-doped sample has less disorder but enhanced moment associated with a B-induced change in the local Yb environment. Because the effective $Yb^{3+}$ moments respond differently to different environments, doping can induce different conduction band response (RRR) and large changes in $Yb^{3+}$ concentration.

The low-temperature magnetic properties of all five samples are dominated by a large upturn in susceptibility we fit to a Curie impurity term below T=150K using, $\chi = \chi_o + C/(T+\Theta)$ where $\chi_o$ is diamagnetic and small (on the order of $1\times10^{-5}$emu/mole). The Weiss temperature varies between 1 K and 5 K, suggesting the existence of weak antiferromagnetic interactions. We estimate the impurity contributions by calculating an effective moment from the Curie constant for each sample. Assuming that this Curie tail arises from the $Yb^{3+}$ ions, we find that the impurity fraction for the alumina slow cooled and yttria slow-cooled samples are 0.040% and 0.035%, respectively while for the alumina-quenched sample it is 0.46%. This very large impurity contribution in the quenched sample suggests that the magnetic properties of samples prepared in this manner are dominated by impurity effects rather than intrinsic magnetic properties. Magnetic susceptibility of the C and B-samples, display increased effective moment when compared to the slow-cooled alumina and yttria-crucible-grown samples. The effective moments are found to correspond to 0.070% $Yb^{3+}$ impurity concentration for the C-doped sample and 0.46% for the B-doped sample, and the B-doped sample did not differ from the quenched samples. We did not observe significant changes in effective moment for C and B doped samples when compared with moments obtained from stoichiometric samples. Our susceptibility measurements, in agreement with previous work,[3-5] provide no evidence for a valence transition in this compound. At higher temperatures (300K and above) the susceptibility is dominated by $\chi_o$ and we never observe any signature of a valence transition, as was initially reported.[1]

Finally, we comment on the magnetic feature shown in the inset to Fig. 4b. The quenched sample shows evidence for an antiferromagnetic phase transition at approximately T=2.5K in an applied field of H=1 kOe. This is somewhat lower than would be predicted from the Weiss temperature, $\Theta$ =5.4K, but consistent with only a very small fraction of ordered impurities. The anti-ferromagnetic transition is also seen in heat-capacity measurements, Fig. 5. This ordering was observed in an earlier experiment [4] and was attributed to a small amount of impurity phase of $Yb_2O_3$. From the entropy we find that the magnetic ordering corresponds to a 4 % concentration of $Yb^{3+}$. The discrepancy between the $Yb^{3+}$ moment concentration calculated from the susceptibility (0.46%) and the specific-heat (4%) may arise from the large field (1T) in the susceptibility measurements blocking ordering. None of the slow-cooled samples show any type of magnetic ordering.

From susceptibility measurements we find that the magnetic behavior of all three samples is not related to a valence transition but rather to $Yb^{3+}$ impurity concentrations. The ground state of Yb is the diamagnetic divalent configuration $Yb^{2+}$ and as shown in previous work[3], it is possible with specific preparation procedures to eliminate the $Yb^{3+}$ impurities.

## IV. CONCLUSIONS

Our measurements show a significant decrease in thermal expansion only after YbGaGe was doped with low level (0.5%) C and B impurities. In stoichiometric samples we observe no major changes in the thermal expansion even in the quenched sample with large effective Yb moment. The C- and B-impurities have little or no effect on the effective moment. Variations in effective moment from sample to sample fall in the range that we observe in stoichiometric samples and we see no correlation between the thermal expansion and the Yb effective moments. The unusual changes in the thermal expansion of YbGaGe resulted only after the insertion of C and B into the lattice. The difference in atomic radii between the Yb and the B and C atoms is substantial so we do not expect C and B atoms to occupy Yb positions in the lattice. On the other hand we do expect C and B substitutions in the Ga and Ge planes and the effect we observe must be caused by these substitutions. We establish that the variation of the a and c axes, and thus the volume, with temperature is extremely sensitive to substitutional disorder, i.e. it requires the presence of B or C impurities perhaps eventually driving this metal to ZVE behavior as doping is increased further, as measured in the original work[1]. Such extreme sensitivity of thermal expansion to doping is unusual in a metallic system, suggesting that thermal expansion characteristics of intermetallics can be tuned by substitutional disorder and that YbGaGe could indeed be a ZVE metal if the correct synthesis recipe is used.

This research proceeded under the auspices of the National Science Foundation, the State of Florida, and the United States Department of Energy. Research carried out in part at the NSLS at BNL is supported by the U.S. DOE (DE-Ac02-98CH10886).

TABLE 1: Sample preparation conditions and resulting physical properties for YbGaGe samples

| **Sample** | **Crucible** | **a-axis (Å)*** | **c-axis (Å)*** | **$\beta$ ($10^{-5}$/K)** | **RRR** | **$\mu_{eff}$ ($\mu_B$)** |
|---|---|---|---|---|---|---|
| YbGaGe | Alumina | 4.19579(1) | 16.7202(1) | 2.72 | 1.9 | 0.06 |
| YbGaGe | Alumina (Q) | 4.19950(7) | 16.73779(7) | 3.21 | 1.7 | 0.23 |
| YbGaGe | Yttria | 4.20047(7) | 16.7226(4) | 3.19 | 2.1 | 0.05 |
| C-(0.5%) | Alumina | 4.1965(1) | 16.7360(6) | 1.92 | 1.9 | 0.08 |
| B-(0.5%0 | Alumina | 4.1974(2) | 16.724(1) | 1.70 | 2.2 | 0.23 |

*at T=300K

## Figure Captions

Fig. 1 A profile fit to the synchrotron X-ray powder diffraction pattern observed for yttria grown stoichiometric sample at 10K. The sets of tick marks below the data indicate the positions of the allowed reflections. The lower curves represent the differences to the observed profiles on the same scale.

Fig. 2. Temperature dependence of the unit-cell volume and edge lengths are plotted (a) for the yttria grown stoichiometric sample; and, (b) the B-doped alumina-grown sample. The lines are a polynomial fit and are meant to be a guide to the eye.

Fig. 3 Temperature dependence of the normalized unit-cell volumes ($V/V_{10K}$ or $V/V_{20K}$ ) for all samples. Lines are polynomial fits and serve as guides to the eye.

Fig. 4 Normalized resistance measurements of single crystalline YbGaGe prepared in alumina or yttria crucibles with different cooling rates (a). The effect of C or B doping of YbGaGe is shown as the RRR values and reflect the amount of disorder in the lattice. Magnetic-susceptibility measurements (*M*/*H*) at $H$ = 1 T of stoichiometric YbGaGe, and a sample that was quenched, is shown in the inset (b). In the quenched sample at low temperature, there is an antiferromagnetic ordering of the $Yb^{+3}$ moments.

Fig. 5. The magnetic specific heat ($C_{mag}$) was obtained by subtracting $C_{phonon}$ + $C_{electron}$ for the same YbGaGe quenched sample shown in the inset of Fig. 1 (b). To quantify the $Yb^{3+}$ moments present, an extrapolation to $T$ = 0 K assumed $C=\beta_3T^3$, with $\beta_3$ evaluated from a fit to the lower six data points. For the high-temperature tail an extrapolation was made using the empirical expression $C_{mag}/T=D_4/T^5+D_3/T^4+D_2/T^3$ and evaluating the parameters by fitting data between ~4 and 6 K. Integration of $C_{mag}/T$ from 0 K to 6 K gives a magnetic change in entropy of $\Delta S_{mag}$ = 180 mJ $K^{-1}$ $mol^{-1}$.

.

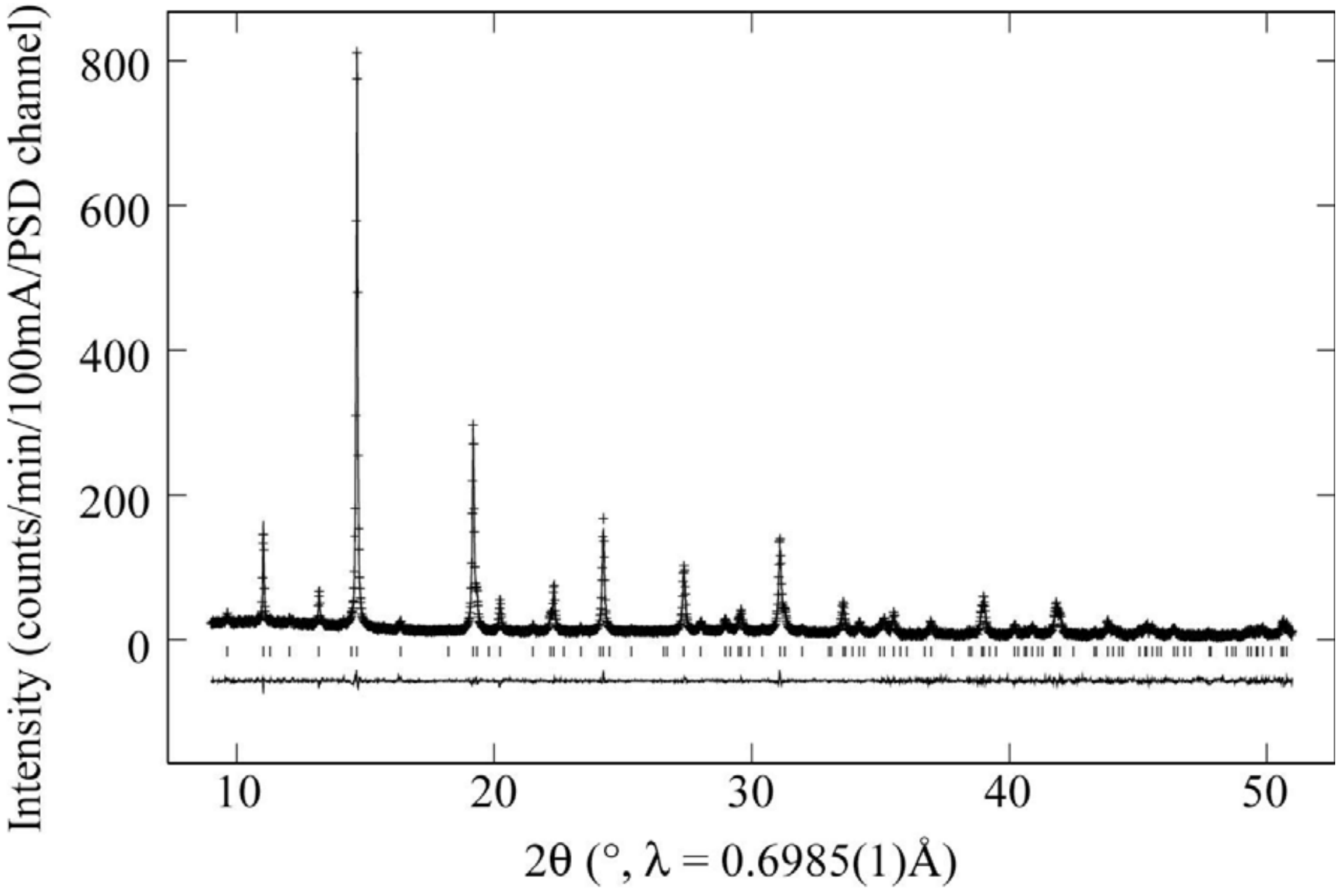

Figure1.

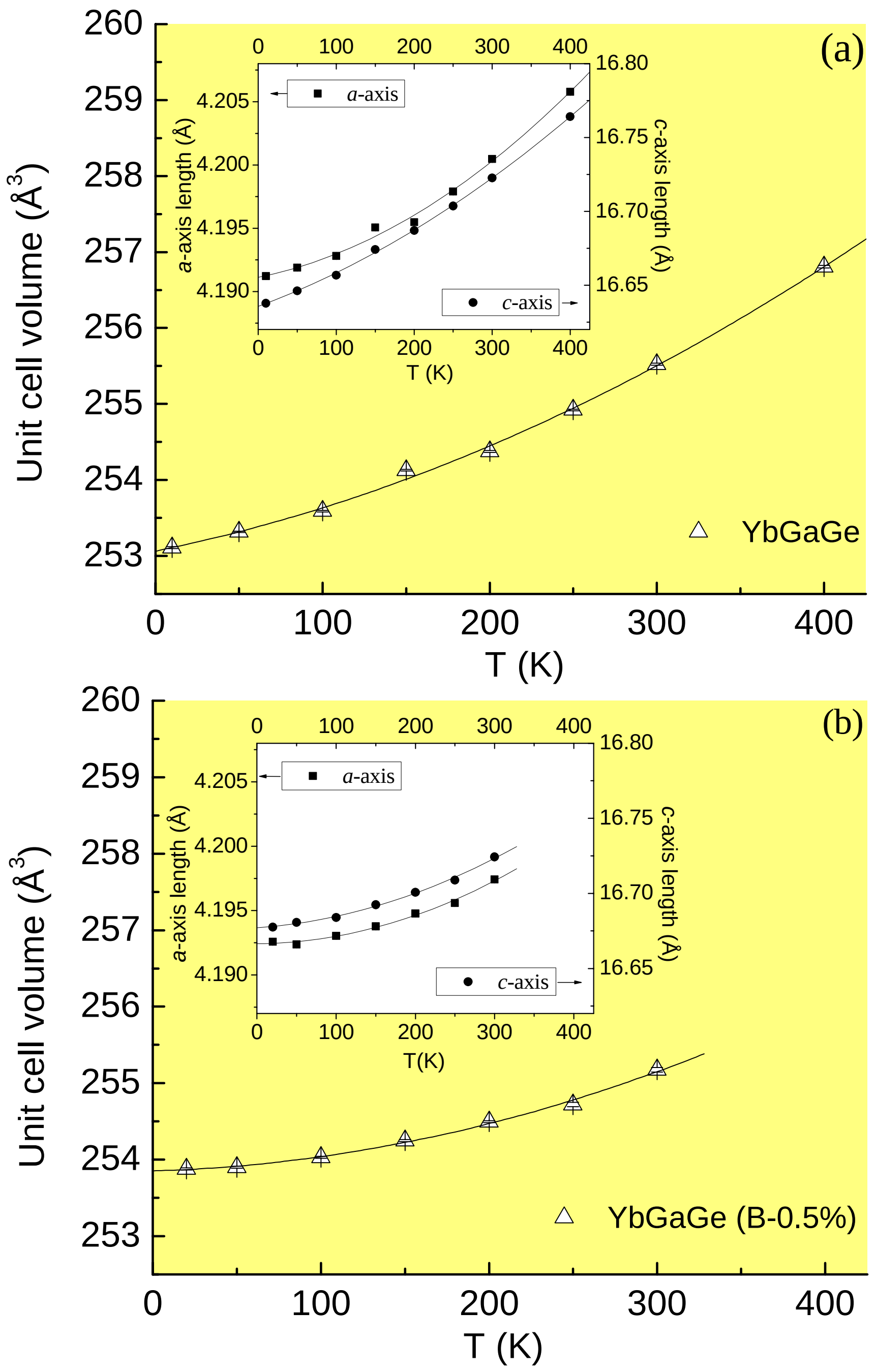

Figure 2

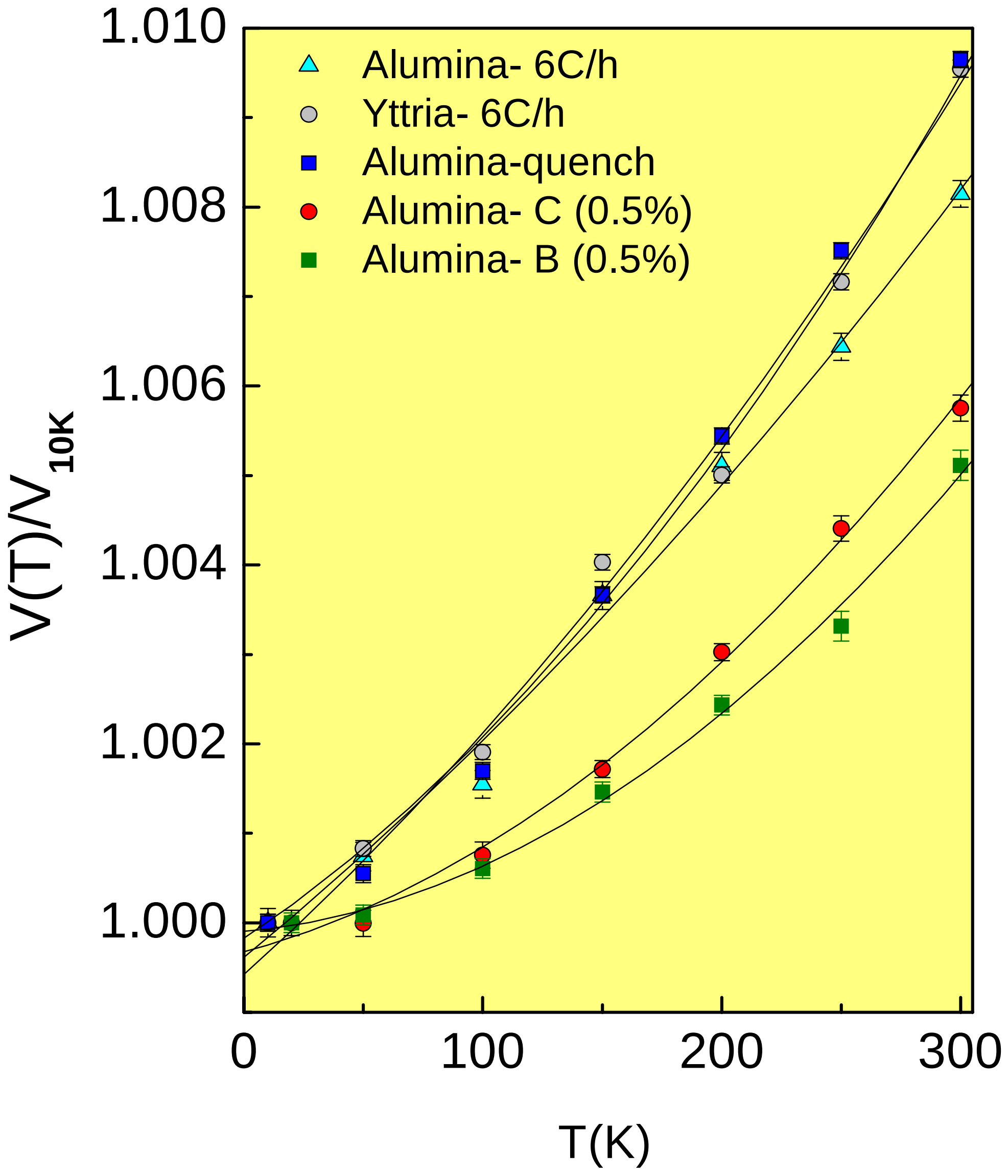

Figure 3

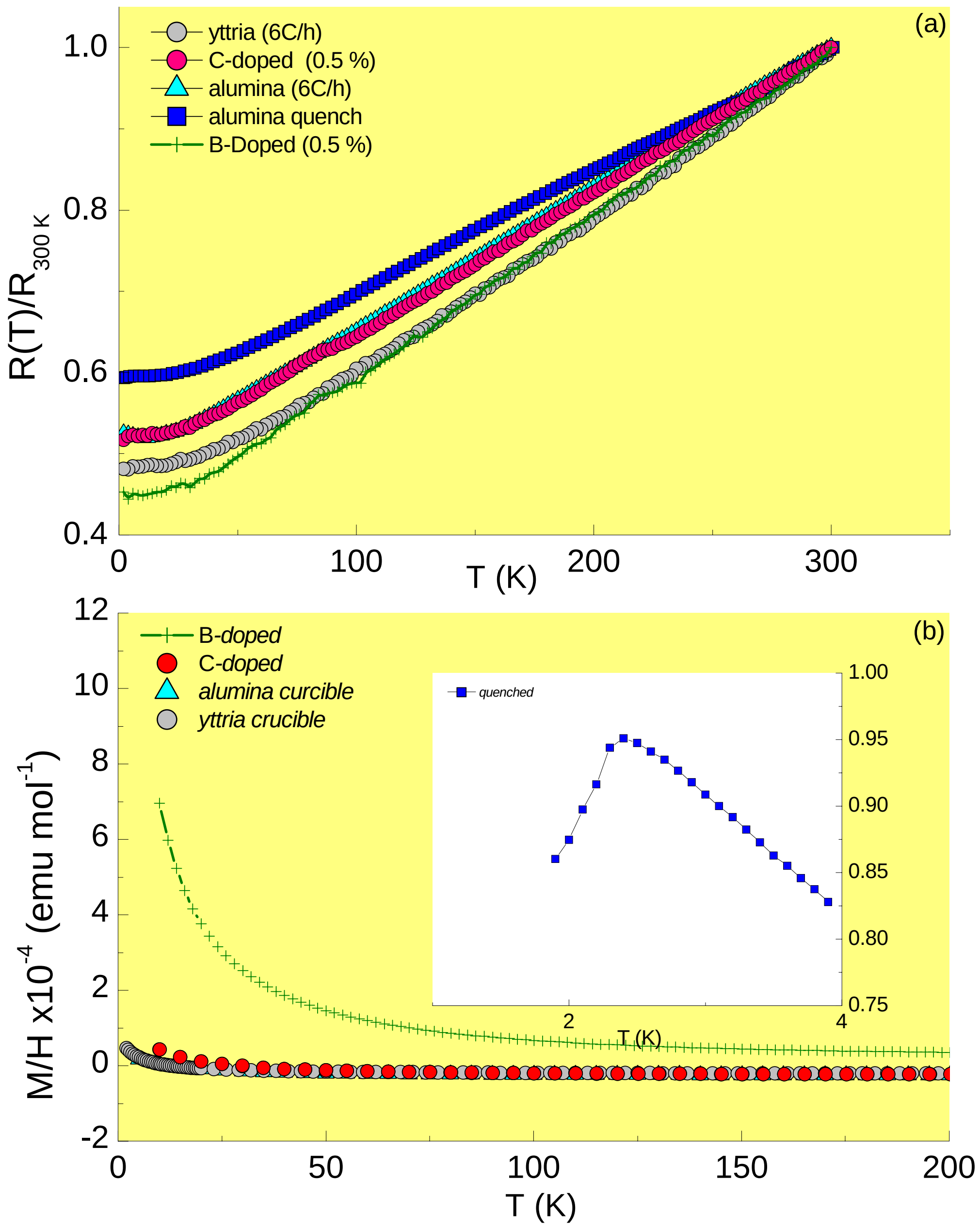

Figure 4

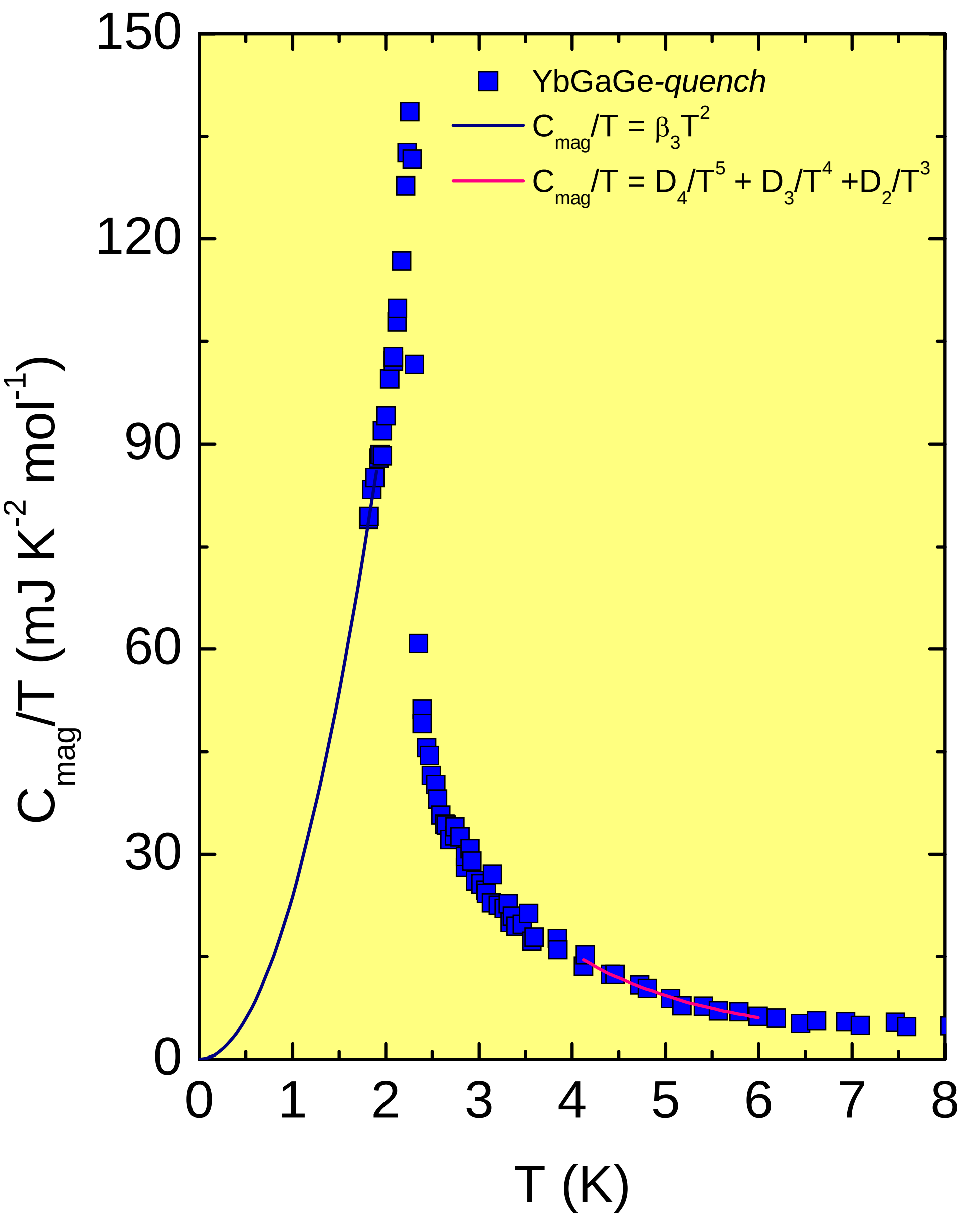

150
120
90
60
30
0
0 1 2 3 4 5 6 7 8
T (K)
$C_{mag}/T$ (mJ $K^{-2}$ $mol^{-1}$)
YbGaGe-*quench*
$C_{mag}/T = \beta_3 T^2$
$C_{mag}/T = D_4/T^5 + D_3/T^4 + D_2/T^3$

Figure 5